*Ab initio* linear response and frozen phonons for the relaxor PMN (PbMg$_{1/3}$Nb$_{2/3}$O$_3$)

Narayani Choudhury[1,2], Zhigang Wu[1], E.J. Walter[3] and R.E. Cohen[1]

[1]Geophysical Laboratory, Carnegie Institution of Washington
5251 Broad Branch Road N.W., Washington, DC 20015
[2]Solid State Physics Division, Bhabha Atomic Research Centre,
Trombay, Mumbai 400 085, India
[3]Department of Physics, College of William and Mary, Williamsburg, VA 23187

**Abstract**

We report first principles density functional studies using plane wave basis sets and pseudopotentials and all electron linear augmented plane wave (LAPW) of the relative stability of various ferroelectric and antiferroelectric supercells of PMN for 1:2 chemical ordering along [111] and [001]. We used linear response with density functional perturbation theory (DFPT) which is implemented in the code ABINIT to compute the Born effective charges, electronic dielectric tensors, long wavelength phonon frequencies and LO-TO splittings. The polar response is different for supercells ordered along [111] and [001]. Several polar phonon modes show significant coupling with the macroscopic electric field giving giant LO-TO splittings. For [111] ordering, a polar transverse optic (TO) mode with E symmetry is found to be unstable in the ferroelectric *P3m1* structure and the ground state is found to be monoclinic. Multiple phonon instabilities of polar modes and their mode couplings provide the pathway for polarization rotation. The Born effective charges in PMN are highly anisotropic and this anisotropy contributes to the observed huge electromechanical coupling in PMN solid solutions.

# I. INTRODUCTION

Lead magnesium niobate $Pb(Mg_{1/3}Nb_{2/3})O_3$ (PMN) is a relaxor having a large, broad and frequency dependent peak in the dielectric constant versus temperature[1] at around 300K. This diffuse behavior in relaxors is unlike ordinary ferroelectrics, which exhibit an abrupt change in the structure and properties at the Curie temperature[2]. PMN is the relaxor end member of the large strain piezoelectric solid solution PMN-PT which is a relaxor ferroelectric below certain $PbTiO_3$ (PT) concentration[2].

PMN has the perovskite structure[3] with the $Mg^{2+}$ and $Nb^{5+}$ cations present in the correct ratio for charge balance but located randomly on the B site which has only short range order. Similar to other relaxors, PMN is heterogeneous with respect to chemical, compositional and polarization order on a nanometer scale, but it is typically cubic on a micrometer scale at room temperature. The effects of this disorder manifest in all physical observations and have therefore spurred extensive experimental studies of the structural heterogeneities[3-19] and phase transitions[3-7], diffuse scattering[5] and spin-glass type behavior of relaxors[20]. The dynamical properties of relaxors reveal anomalies in their observed specific heat[11], high pressure Raman[8] and inelastic neutron spectra[12-14]. The degree of order observed experimentally depends on the length and time scale of the experimental probe. Detailed theoretical first principles studies to understand and analyze the dynamical behavior and their relationship to the observed properties of these systems are therefore desirable.

Due to the complexity of relaxors, theoretical studies necessarily have to adopt a multiscale approach using several techniques to cover the length and time scales involved. As a first step, we have performed first principles all electron LAPW and linear response studies using the code ABINIT[21] for ordered supercells with 1:2 ordering along the [111] and [001] directions in PMN. Whereas the first principles calculations yield some insight on the behavior of ordered PMN, real PMN is a disordered solid solution on average. Studies of ordered supercells are important for developing interatomic potentials[22] and effective Hamiltonians[23], required to study finite temperature effects of relaxor ferroelectrics as well as their behavior over larger length and time scales. Using techniques like molecular beam epitaxy, one can in principle make ordered PMN, which would be ferroelectric according to our results. Thus our calculations are also predictions for the behavior of ordered PMN, which might someday be synthesized.

The goals of the present study are (i) to determine the role of the crystal structure and bonding in promoting ferroelectric relaxor and enhancing piezoelectric behavior and to understand the role of phonon instabilities, and (ii) to understand the factors that lead to piezoelectric enhancement with polarization rotation. LAPW calculations involving several ordered supercells of PMN were used to explore the energy landscape to understand the relative energies of possible ferroelectric (FE) and antiferroelectric (AFE) states. Norm conserving pseudopotentials were developed for first principles studies of the structural and vibrational properties, and comparisons of the energy landscape with LAPW computations helped develop accurate pseudopotentials required for the linear response studies.

First principles theoretical calculations[24-37] using all electron LAPW[24-29] as well as plane wave basis sets and pseudopotentials[30-35] have been successfully used to understand fundamental aspects related to crystal structure, bonding and ferroelectric properties[24], spontaneous polarization and piezoelectric constants[25-27,37], polarization rotation and observed high electromechanical coupling coefficients[25,26], frozen phonon[29-31] and linear response studies[28,32,33], structural response to macroscopic electric fields[36], *etc.* for a wide variety of ferroelectric materials[23]. First principles frozen phonon studies for various ordered supercells in PMN have been reported previously by Prosandeev *et al.*[30,31] using ultrasoft pseudopotentials. We have built on the previous work by performing first-principles lattice dynamics computations including LO-TO splittings, and have studied a variety of different structures, following a cascading set of soft modes towards the ground state.

PMN-PT attracts interest as a giant piezoelectric single crystal material with enhanced piezoelectric and electromechanical coupling coefficients and finds important applications[2] in piezoelectric transducers and actuators, non-volatile ferroelectric memories, pyroelectric arrays, dielectrics for microelectronics and wireless communication, non-linear optical applications, etc. Here we report first principles density functional ground state and linear response studies of phonon instabilities, anisotropic Born effective charges, electronic dielectric tensors, long wavelength phonon frequencies and LO-TO splittings for various ordered supercells of PMN.

Sections II and III give details about the techniques adopted and results obtained respectively, while section IV gives the summary and conclusions.

## II. TECHNIQUES

### A. LAPW studies

The calculations were performed using the first-principles all-electron linearized augmented plane-wave (LAPW) method with the local-orbital extension[38]. Core states are treated fully relativistically. The Hedin-Lundqvist local-density approximation was used for the exchange-correlation energy. The muffin-tin radii are 2.1, 1.7, 1.7, and 1.6 Bohr for Pb, Mg, Nb and O respectively. The LAPW convergence parameter $RK_{max}$ = 7.5, corresponding to about 2400 LAPWs, was found to be well converged. The special k-point method with a 4×4×4 mesh was used to sample the Brillouin zone; selected computations with a 6×6×6 mesh showed that the former mesh was sufficient.

### B. Linear response studies

The linear response computations used the density functional perturbation theory (DFPT)[21,39-44] with a plane-wave basis and norm conserving pseudopotentials as implemented in the code ABINIT[21] within the local density approximation for electron exchange and correlation. We generated norm conserving pseudopotentials (Table I) using the code OPIUM[45-47]. The Pb ($5d$, $6s$ and $6p$), Mg ($2s$, $2p$ and $3d$), Nb ($4s$, $4p$ and $4d$) and the O ($2s$ and $2p$) were treated as valence and the parameters used are listed in Table I. A kinetic energy cutoff of 100 Ryd was used and the Brillouin zone (BZ) integration were performed using a 4×4×4 k-point mesh.

## III. RESULTS

### A. Structural relaxation

The structures of the ordered supercells (Tables II-VI) used in the present studies are shown in Fig. 1. The ideal positions for the 15-atom cell with 1:2 ordering along the [111] have hexagonal ($P\bar{3}m1$) symmetry. All our calculations were done at the LDA zero pressure volume (1304 Bohr$^3$ for the $P\bar{3}m1$ structure). Symmetry preserving structural relaxation starting from the ideal positions in the $P\bar{3}m1$ structure leads to a non-polar antiferroelectric (AFE) $P\bar{3}m1$ structure (Tables II, III). Polar displacements along the softest mode in the $P\bar{3}m1$ AFE structure along [111] lead to a polar $P3m1$ symmetry. Using this as a starting structure, we relaxed the atomic positions and cell parameters at fixed volume and the relaxed structure obtained with $P3m1$ symmetry is given in Table IV. The relaxed FE and AFE structures (Tables II-IV) obtained using the LAPW and pseudopotential techniques are found to be in good agreement.

The polar ferroelectric $P3m1$ structure (Table IV) has a lower energy than the non-polar antiferroelectric $P\bar{3}m1$ structure (Table III), with an energy difference of 12 mRyd/15-atom. The AFE to FE double well obtained from LAPW and pseudopotential calculation are compared in Fig. 2. The difference in energy between the AFE and FE provides a stringent test and significant effort was required for deriving accurate pseudopotentials that yield results in agreement with these LAPW results. The AFE structure in Fig. 2 corresponds to a maximum in energy and the associated lattice instability drives it to the lower energy FE $P3m1$ structure.

The $P3m1$ FE structure has a doubly degenerate unstable phonon mode. Displacements of atoms using a linear combination of the doubly degenerate unstable E symmetry phonon mode eigenvector yields the monoclinic $Cm$ (E) structure which has a slightly lower energy (~ 1.1 mRyd/15-atom) than the $P3m1$ structure. Symmetry preserving structural relaxation of the $Cm$ (E) structure yields the lower energy $Cm$

(R) structure having an energy difference of 6.7 mRyd/15-atom with respect to the *P3m1* FE structure. (Table II). The structural differences between the *Cm* (E) and *Cm* (R) structures are that the displacement pattern involving the *P3m1* to *Cm* (E) structure (obtained from the unstable phonon eigenvectors) does not involve any displacement of the Pb, Mg and Nb atoms along the hexagonal *c*-axis (cubic [111]). It is possible to have a lower energy monoclinic *Cm* (R) structure, which allows displacement of the Pb, Mg and Nb along the hexagonal *c*-axis. The computed *Cm* (E) and *Cm* (R) structures are shown in Table VI. The *Cm* (E) and *Cm* (R) monoclinic structures have an A" symmetry unstable phonon and the ground state is found to be monoclinic (*C2*) having an energy difference of 2.69 mRy/15-atom with respect to the *Cm* (R) structure. Structural relaxation of PMN at the experimental volume (1343 Bohr$^3$) also yields similar results. Prosandeev *et al.*[30,31] found a triclinic ground state when relaxing a 30 atom ordered supercell of PMN.

The ideal positions with 1:2 ordering along the [001] corresponds to tetragonal symmetry (*P4/mmm*) (Table V). Symmetry preserving structural relaxation give an antiferroelectric *P4/mmm* structure. Although, an FE *P4mm* structure (involving symmetry lowering displacements of atoms involving loss of inversion symmetry) possibly exists, symmetry preserving structural relaxations starting from the FE *P4mm* structure do give the higher AFE symmetry with the Pb, Mg, Nb and O atoms moving to Wyckoff positions in the *P4/mmm* symmetry (Table V). We were unable to find a stable polar [001] ordered structure with polarization along [001]. However, for displacements along [100], a polar orthorhombic FE structure (space group *Pmm2*) (Table II) was obtained.

Experimental x-ray diffuse scattering studies of antiferroelectric fluctuations in PMN[5] have been reported; these however involve octahedral tilts. Some effort to explain the origin of relaxor behavior based on these observed results have been attempted[5]. More systematic theoretical studies with larger cells, volume, pressure and temperature dependence, including effects of disorder etc. are required for comparison with these experiments on real disordered PMN. It is likely that by having supercells with larger number of atoms, other antiferroelectric phases involving rotation of the $MgO_6$ and $NbO_6$ octahedral as seen in $PbZrO_3$[29] may also be obtained; such phases are hindered in the present calculations where some atoms are constrained to be fixed by symmetry and the periodic boundary condition.

Our computed relaxed structures and relative energies are in agreement with the results of Prosandeev *et al.*[30] for the *P4/mmm* and $P\bar{3}m1$ supercells. Detailed comparisons however are difficult as their computations are at different volumes. Their studies[30] also suggest that the total energies are lower for 1:2 chemical ordering along the [111] as compared to the [001], in agreement with our results.

**B. Born effective charges**

The Born effective charge tensor gives a measure of the local dipole moment which develops when the nuclei are moved and corresponds to the variation of the polarization with atomic displacements. As the Born effective charge tensor $Z^*_{\alpha\beta}(k)$ is a mixed second derivative of the total energy, with respect to macroscopic electric field component $E_\alpha$ and atomic displacement component $\tau_{k,\beta}$. There is no requirement that the tensor be symmetric[28].

The computed components of the Born effective charges are completely symmetric for all the atoms with all non-diagonal elements vanishing in the tetragonal *P4/mmm* AFE structure, where the Cartesian axes coincide with the principal axes directions. For the hexagonal [111]FE and AFE structures, the Born effective charges in the Cartesian frame have (i) small non-vanishing off-diagonal elements, and (ii) the tensor is symmetric for the Pb, Mg and Nb atoms and (iii) the tensor is not symmetric for the oxygen atoms for the *P3m1* and $P\bar{3}m1$ symmetry. For the hexagonal [111] FE and AFE structures, we have obtained the principal values of the charge tensors and the principal axes (Tables VII-X). The Cartesian Born effective charges for the *P4/mmm* tetragonal[30,37] and hexagonal[30] $P\bar{3}m1$ symmetry have been reported previously, and are in good agreement with our results. These are the first studies of the principal axes contributions.

The average changes in charge tensors between the *P3m1* FE and $P\bar{3}m1$ AFE structures for the Pb, Mg, Nb and O are small. The computed charges are consistent with the values typically reported for other perovskites[33]. The effective charges of Pb, Nb and O are significantly larger than their nominal ionic values (2, 5, and -2, respectively) which has been found to be a universal feature of nearly all of the ferroelectric perovskites[23]. The origin of this behavior has been traced to the ionic-covalent character and specifically the hybridization between O $2p$ and B-atom $3d$ or $4d$ orbitals in the $ABO_3$ type materials[23,28].

Both the hexagonal and tetragonal systems are uniaxial systems, having the hexagonal or tetragonal axis as the axis of symmetry. The principal axis values of the Born tensors for the various atoms for the *P3m1*, *P$\bar{3}$m1* and *P4/mmm* supercells have similar symmetry with only two independent diagonal elements for the Pb, Mg and Nb, and three unequal diagonal elements for the oxygen atoms. The principal axes of all the cations are along the hexagonal **a** and **c** axes in the hexagonal FE and AFE structures. The major principal axes of all the oxygen atoms are along the Nb-O bond directions in the tetragonal (hexagonal) supercells. The magnitude of the Born effective charges of O atoms are significantly larger along the Nb-O bonds than in other directions and this anisotropy is due to the covalency of the Nb-O bonds.

The principal values of the Born effective charge tensors for the monoclinic *Cm* (R) structure are given in Table X. All the cations and the oxygens (O2, O6, and O7) in Wyckoff position *1a*, have the monoclinic unique axis along Cartesian (1,-1, 0) as a principal axis. The direction of spontaneous polarization is also a principal axis for some of the cations and oxygens. The principal values of charge tensors are quite similar in the *Cm* (R) and [111] FE *P3m1* structure, the orientation of their principal axes are however different and this brings about important changes in their spontaneous polarization and piezoelectric properties.

The charge tensors as well as their principal values are significantly different for the [111] ordered supercell and the [001] ordered tetragonal supercells. Although the Mg atoms have charge values close to their nominal ionic values for the [001] supercells (as predicted earlier using supercell calculations[37]), they are significantly enhanced in the [111] ordered supercells. On the other hand the Nb atoms have larger charges for ordering along [001]. These changes are also reflected in the oxygen atoms and these differences bring about important changes in their polarization, polar response and vibrational characteristics. The computed Born effective charge tensors (Table IX) are found to be in excellent agreement with the reported values for the *P4/mmm* tetragonal[30,37] and *P$\bar{3}$m1* hexagonal[30] symmetries; there are no earlier studies of the FE *P3m1* and monoclinic *Cm* structures.

In Fig. 3, we display the Born effective charge tensors of the [111] ordered supercells as charge ellipsoids. Quadric representation[48] as ellipsoids is generally possible for symmetric second rank tensors. While the Born effective charge tensor is completely symmetric for the Pb, Mg and Nb atoms in the hexagonal structure, it is not symmetric for the oxygen atoms in the *P3m1* and *P$\bar{3}$m1* structures. We first resolved the tensor into an isotropic, symmetric and a traceless antisymmetric part as described in Ref. [49]. It was verified that the symmetrized charge tensor gives the LO and TO frequencies (Fig. 4) correctly within 4 cm$^{-1}$, and the charge tensor was symmetrized (for the oxygens atoms only) for display. The computed charge tensors were converted from Cartesian to hexagonal coordinates and then suitably scaled to give ellipsoids appropriate for typical values of mean square atomic displacements. The ellipsoids were rotated appropriately for the equivalent atoms according to the crystal symmetry, and displayed in Fig. 3 using the software *xtaldraw*[50].

In Fig. 3 we compare the charge tensors in the hexagonal *P3m1* and *P$\bar{3}$m1* structures viewed along and perpendicular to the direction of polarization. The macroscopic polarization direction in the [111] FE structure is a principal axis for the Pb, Mg and Nb atoms but not for the oxygen atoms. The sum over equivalent atoms must be consistent with the crystal point group symmetry, so fully symmetric phonon modes have no enhancement of the polarization from the effective charge asymmetries. However, for less symmetric modes or under strain, the anisotropies promote enhancement of the piezoelectric response oblique to the [111] axis. Thus, rotation of polarization towards directions which are along the major principal values and which maximize contributions from the cations and anions with corresponding ferroelectric displacements can give rise to significant enhancement of piezoelectric properties. As the anisotropy of the charges of the cations are much smaller than the corresponding values for oxygen, significant enhancement can be expected along the major principal oxygen axes (along pseudocubic directions), particularly in the basal plane where the majority of the cations also have major principal axis contributions. Maximum enhancement can be expected by alignment of the Nb and the oxygen atoms, owing to their large Born effective charge values.

**C. Electronic dielectric tensor**

Both the electronic dielectric tensor and Born effective charges are required to understand the influence of the macroscopic electric field on phonon properties. We have studied the variations of the high-frequency electronic dielectric tensor for the various ordered supercells in PMN. For both the hexagonal and tetragonal systems[48] there are only two independent diagonal components: parallel and perpendicular to the hexagonal (tetragonal) symmetry, with all non-diagonal elements vanishing.

The electronic dielectric tensor has $\varepsilon_{11}$ = 6.79, and $\varepsilon_{33}$ = 6.43 in the *P3m1* structure and their corresponding values are 7.24, 6.91 in the $P\bar{3}m1$ structure. The electronic dielectric tensor for the tetragonal symmetry has $\varepsilon_{11}$ = 7.23, $\varepsilon_{33}$ = 7.08. The electronic dielectric tensor in the monoclinic *Cm* (R) structure has principal values of 6.40, 6.42 and 6.63 with their principal axes along the Cartesian (-1,-1,0), (0, 0,1) and (1, -1, 0) directions, respectively. The computed average electronic dielectric constant in the *P3m1* and *Cm* (R) structures is overestimated as compared with experimental[30] infrared observations of around 5.83. These can however be understood as the LDA generally overestimates the dielectric constant. Nevertheless, the LDA results are still expected to be qualitatively correct with regards the relative variations of the dielectric constants for different ordering schemes along various directions, making such studies useful for material design.

**D. Symmetry analysis of phonon modes: Group theoretical details**

Group theoretical analysis[51] was undertaken to derive the symmetry vectors for classification of the zone center phonon modes into their irreducible representations for 1:2 chemically ordered supercells of PMN (Tables III-V) having space group *P3m1*, $P\bar{3}m1$, *Cm* and *P4/mmm*. This analysis will help in determining Raman and infrared contributions from ordered regions to the observed spectra. Although all of these structures are saddle points on the energy surface, they can be stabilized by finite temperature and anharmonicity. Anharmonicity will break the selection rules, but it is still useful to understand the underlying symmetry of the harmonic contributions.

**1. [111] FE *P3m1* structure**
The phonon modes at the Γ point for the ferroelectric *P3m1* structure with 1:2 ordering along [111] can be classified as

Γ: 12 $A_1$ + 3$A_2$ +15E

The polar $A_1$ and doubly degenerate E modes involve LO-TO splittings and 39 of the total 42 optic modes in this symmetry are polar. The $A_1$ and E phonon modes are both Raman and IR active, while the $A_2$ modes are optically inactive but can be measured from inelastic neutron scattering. The $A_1$ modes involve in-phase displacements along the hexagonal *c*-axis, while the doubly degenerate polar E modes involve displacements along the hexagonal *a* and *b* directions. By symmetry, the displacements on the Pb, Mg and Nb atoms vanish and the $A_2$ modes involve only out-of-phase displacements of the oxygen atoms. The decompositions of the group theoretical irreducible representations (IRD) for the given Wyckoff position at the Γ point are shown in Tables III-V, for these various supercells.

**2. [111] AFE $P\bar{3}m1$ structure**

For the AFE [111] $P\bar{3}m1$ structure, the phonon modes at the zone center can be classified as

Γ: 4$A_{1g}$ +2$A_{1u}$ + $A_{2g}$ + 8$A_{2u}$ + 5 $E_g$ + 10$E_u$

The higher symmetry of the AFE structure increases the number of irreducible phonon representations. Inspecting the character tables for these irreducible representations, one can write the compatibility relations between the phonon modes in the FE and AFE structures. These give the following correspondences

$12A_1$ (*P3m1*) → $4A_{1g}+8A_{2u}$ ($P\bar{3}m1$)
$3A_2$ (*P3m1*) → $A_{2g}+2A_{1u}$ ($P\bar{3}m1$)
$15E$ (*P3m1*) → $5E_g+10E_u$ ($P\bar{3}m1$)

The $A_{1g}$ and $E_g$ modes are Raman active, and the $A_{2u}$ and $E_u$ phonons involve polar infrared modes, while the $A_{2g}$ and $A_{1u}$ modes are optically inactive. The decompositions of the group theoretical irreducible representation (Table III) reveal that there is no Raman activity from the Pb atoms in *1a*, Mg *1b* and O *3e*, and the optically inactive modes involve only the vibrations of the oxygen atoms. The $A_{2u}$ and $E_u$ phonons involve polar modes and the $E_g$ and $E_u$ modes are doubly degenerate.

**3. [111] FE *Cm* structure**

Due to the low symmetry of the monoclinic *Cm* structure, there are only two representations A' and A" at the zone center, and the phonon modes can be classified as

Γ: $27A' + 18A''$

All the zone center optic phonon modes are Raman and IR active.

**4. [001] AFE *P4/mmm* structure**

The phonon modes at the Γ point can be classified as

Γ: $4A_{1g}+8A_{2u}+B_{1g}+2B_{2u}+5E_g+10E_u$

The $A_{2u}$ and doubly degenerate $E_u$ phonon modes are polar infrared active, and the $A_{1g}$ and $E_g$ modes are Raman active, while the $B_{1g}$ and $B_{2u}$ modes are non-polar and optically inactive. While the $A_{2u}$ polar modes involve in-phase displacements of atoms along the *z*-axis, the $E_u$ modes involve displacements along the tetragonal *a* and *b* axis. By symmetry, the $B_{1g}$ and $B_{2u}$ modes have no displacements of all the Pb, Mg and Nb atoms, and involve only out-of-phase displacements of the oxygen atoms along the *z*-direction.

**E. Phonon frequencies**

The computed long wavelength phonon frequencies for the [111] FE *P3m1*, [111] FE *Cm* (R), [111] AFE $P\bar{3}m1$ and [001] AFE *P4/mmm* supercells are compared with available Raman[8,30] and infrared[30] data in Fig. 4. In Fig. 4(a) we show the unassigned zone center TO modes and the Raman active and IR active modes with their symmetry assignments. The corresponding LO frequencies in the *P3m1* structure are shown in Fig. 4(b). There is a single unstable phonon mode having A" symmetry in the monoclinic *Cm* (R) and a doubly degenerate unstable E mode in the *P3m1* structure, while the other supercells have multiple phonon instabilities. The phonon spectral ranges in these supercells are found to be in good agreement with the available Raman[8,30], infrared[30] and inelastic neutron data[15], as well as the phonon frequencies obtained from frozen phonon calculations on PMN[30]. The phonon frequencies of PMN *Cm* (R), *P3m1* and $P\bar{3}m1$ structures using supercells with 1:2 ordering along [111] are quite similar, although the FE *Cm* (R) and *P3m1* structures are significantly more stable than the $P\bar{3}m1$ AFE structure. Although the spectra with ordering along [111] and [001] are similar, there are important differences in the long wavelength frequency distributions and polar behaviors which correlate well with the differences in their Born effective charge tensors.

The comparison between the computed frequencies in the *P3m1* FE structure and experimental long wavelength frequencies is satisfactory, especially as these are ordered supercells, whereas real PMN is disordered. Except for the observed mode at 88 cm$^{-1}$, the other phonon modes including the LO and TO frequencies[30] are even in good quantitative agreement with experiments. In the IR experiments[30], the observed low frequency 88 cm$^{-1}$ TO mode is a soft mode with very strong temperature dependence. The inelastic neutron powder data[15] and the observed infrared[30] and Raman spectra[9,30] show low frequency phonon peaks around 30-50 cm$^{-1}$, which are also obtained in the *P3m1* and *Cm* (R) FE structures. These low frequency spectra can contribute to the observed anomalies in low temperature specific heat[11].

**F. Phonon instabilities and polarization rotation**

The doubly degenerate E mode eigenvectors of the unstable phonon mode in the *P3m1* structure involve displacements of all the cations along the hexagonal **a**-axis (or along **b**, having degenerate frequency) and displacements of oxygens consistent with this symmetry. By using different linear combinations of these eigenvectors, one can rotate the polarization in the basal plane. The *Cm* (R) structure have superposed on these E type displacements, additional displacements along the hexagonal *c*-axis of the cations having an $A_1$ phonon mode character. This coupled E-$A_1$ phonon mode again rotates the polarization. The combination of these three modes can give monoclinic or triclinic symmetry.

For the *P3m1* structure, we computed the spontaneous polarization using the Berry's phase[52] method with a 4×4×16 k-point mesh. The computed spontaneous polarization along the ferroelectric direction (Cartesian [111]) in the *P3m1* FE structure is 0.47 C/m$^2$ (LAPW) and 0.52 C/m$^2$ (ABINIT), and the two values are in good agreement. The polarizations of the monoclinic *Cm* structures were evaluated from the computed Born effective charges and the atom displacements from ideal positions. The *Cm* (E) structure has a polarization $P_x = P_y = -0.43$ C/m$^2$, and $P_z = -0.095$ C/m$^2$, which gives a polarization magnitude of 0.62 C/m$^2$ at an angle 27° with the cubic [111] direction. The polarization of the relaxed *Cm* (R) structure yields $P_x = P_y = -0.33$ C/m$^2$, and $P_z = 0.45$ C/m$^2$, which gives a polarization magnitude of 0.65 C/m$^2$ at an angle of 79° with the direction of spontaneous polarization (along [111]) for the *P3m1* structure.

In the *P3m1* [111] FE structure, the polar transverse modes have E and $A_1$ type symmetries. The $E_u$ and $A_{2u}$ phonon modes in the tetragonal *P4/mmm* supercell for ordering along [001] similarly involve polar displacements perpendicular to and along the tetragonal axis respectively, and unstable phonon modes with these symmetries can rotate the polarization.

We show the frozen phonon surfaces in Fig. 2 for successive displacements starting from the AFE structure to the FE structure. The lattice instability drives the $P\bar{3}m1$ to *P3m1* transition. The *P3m1* structure has an E type unstable mode and its eigen displacements involve a phase transition from the *P3m1* to the *Cm* (E) structure along with polarization rotation. These multiple phonon instabilities and their associated phonon mode couplings provide the pathway for polarization rotation.

For the [001] ordering we found no polar state with polarization along [001], but we did find an orthorhombic *Pmm2* state with [100] ordering having a polarization along [001] with value 0.503 C/m$^2$. In this orthorhombic structure, the polarization is perpendicular to the chemical ordering direction.

**G. Long wavelength frequency distribution**

To understand the differences in vibrational properties between the [111] and [001] ordered supercells, we computed the long wavelength frequency distributions (Fig. 5) for analyzing the dynamical contributions from various atoms. These are conceptually similar to the total and partial phonon density of states[53,54], but do not include effects of phonon dispersion. In the present studies, which involve only long wavelength excitations, the corresponding equations for total and partial frequency distributions do not involve the wavevector related Brillouin zone summation in the definition of the phonon density of states.

In Fig. 5, the unstable phonon modes (with imaginary frequencies) are indicated as negative frequencies, and the peak at zero frequency is from the acoustic phonon modes. Contributions from TO and LO modes have been included in Fig. 5. The spectra have been broadened using Gaussians with full width at half maximum of 8 cm$^{-1}$. It is noticed that: i) Despite the complex structure of PMN, the vibrational spectra of the cations are clearly well separated and span distinct spectral ranges. The Pb atoms contribute in the 0-200 cm$^{-1}$ range, and the Nb atoms contribute in the 150-300 cm$^{-1}$ range. The Mg atoms have a very sharp peak around 320 cm$^{-1}$ in the [111] FE and AFE structures but get spread out in the tetragonal structure. ii) The unstable phonon modes in the [111] ordered supercells involve mainly the Pb-O vibrations, while the unstable modes of the [001] structure involve the Nb and O atoms. iii) The vibrational spectra and relative contributions from various atoms in the [111] ordered supercells with *P3m1*, *Cm* (R) and *P$\bar{3}$m1* structures are quite similar. The relative changes in the frequency distribution of the [111] FE *P3m1* and [111] AFE *P$\bar{3}$m1* structures arise mainly from the softening of the unstable phonon modes in the AFE involving Pb and O vibrations. In the monoclinic *Cm* (R) structure, the E phonon mode degeneracies in the *P3m1* structure are lifted and the spectra have broadened. iv) The Mg atom is more ionic in the [001] structure with higher high frequency Mg vibrational spectra, which comes from the shorter ionic bonds. v) The frequency distribution for the [111] and [001] ordered supercells are quite distinct for all the atoms, and these differences largely bring out the corresponding changes in their Born effective charges giving rise to significant changes in their bonding and associated vibrational spectra.

**H. Polar response**

There are several polar modes in PMN with very large mode effective charges (Tables XI-XIV), and these modes involve significant coupling with the macroscopic field. To determine the correspondence between the LO and TO modes, we computed the overlap matrix[33,55,56] between the LO and TO eigenvectors. To understand the LO-TO splitting better, we also compare the LO frequencies computed using DFPT (LO1) with those obtained from oscillator strengths (Fig. 6) and the TO frequencies[43,44,57,58]. The latter method (LO2) neglects the different eigenvector mixing for LO and TO modes. The difference between the LO1 and LO2 frequencies is a measure of the eigenvector remixing.

Zhong *et al.*[33] studied several cubic perovskites and found that the softest unstable phonon modes correlated with the largest LO modes give giant LO-TO splitting. While for simple perovskites, the soft mode is found to have the largest mode effective charge[33], it is interesting to see that in PMN several polar modes have large mode effective charges, some even larger than the soft mode for 1:2 ordering along [111] (Table XI-XIV). Many phonon modes show significant coupling with macroscopic field in PMN and these give rise to giant LO-TO splittings, similar to that obtained for cubic perovskites[33].

The IR oscillator strengths for the [001] and [111] ordered supercells are compared in Fig. 6. The 156i $E_u$ TO unstable phonon mode in the [001] *P4/mmm* tetragonal supercell has the largest oscillator strength (note the break in scale, Fig. 6(a)) and this mode which modulates the covalent bonding of the Nb and O produces the largest polar response. The TO and LO frequencies of this mode are 156i and 665 cm$^{-1}$ respectively, giving giant LO-TO splitting similar to cubic perovskites. The relevant LO-TO splittings are those corresponding to structurally stable states; the present studies however are still useful to understand (i) the large changes in the polar behavior and resultant dielectric enhancement due to the phonon instabilities, and (ii) the variations of polar behavior with polarization rotation.

The 238 cm$^{-1}$ $A_{2u}$ mode in the *P4/mmm* tetragonal structure with large mode effective charge also involves the Nb and oxygen vibrations. The polar strength of Nb is increased in the tetragonal *P4/mmm* supercell due to larger values of the Born effective charge tensor and the mode effective charges. While in the [001] ordered supercell the 156i unstable phonon mode involves the B site Nb cations, the 45i unstable E mode in the [111] FE *P3m1* structure involves the A site (Pb) cations. The LO-TO splittings are distinct for ordering along [001] and [111] (Tables XI-XIV).

Even in the [111] FE structures the largest E and $A_1$ oscillator strength and mode effective charges involve the Nb atoms. In the *P3m1* structure, the 180 cm$^{-1}$ E mode and 259 cm$^{-1}$ $A_1$ mode have larger mode effective charges and IR oscillator strengths (Fig. 6) than the unstable 45i soft mode involving Pb vibrations. The 324 cm$^{-1}$ Mg mode has moderately polar Mg atoms with a mode effective charge of 3.43. The higher frequency strongly polar modes have mainly contributions from the oxygens. The large $A_1$ mode effective charges at 259 and 254 cm$^{-1}$ involve the Nb vibrations. The 121 cm$^{-1}$ mode involve Pb, Nb and O vibrations, while the lower frequencies 69 and 66 cm$^{-1}$ have mainly contributions from the Pb atoms.

The E (TO) mode at 49 cm$^{-1}$ in the *P3m1* FE structure with a mode effective charge of 5.58 is mainly from the Pb atoms in Wyckoff positions *1b*, and this phonon mode is stable unlike the unstable TO 45i cm$^{-1}$ E mode involving vibrations of Pb in Wyckoff positions *1a* and *1c*. The anisotropy of the charge tensor for this atom with larger principal values along the polarization direction is different from that for the other inequivalent Pb atoms, and this manifests in their stability and corresponding LO-TO splittings. Subtle differences in the polar strengths of the crystallographically inequivalent Nb and O atoms are also noticed. As their average Born effective charges are quite similar, it is the difference in displacement pattern that changes their polar strength.

It is interesting to see how the modes harden going from the relaxed [111] AFE $P\bar{3}m1$ structure to the relaxed [111] FE *P3m1* structure (Fig. 6). The Pb vibrational 110i ($E_u$) and 90i ($A_{2u}$) modes in the [111] AFE structure correspond respectively to the 45i (E) and 121 ($A_1$) cm$^{-1}$ modes in the [111] FE *P3m1* structure. Similarly, the 180 (E) and 259 ($A_1$) cm$^{-1}$ phonon modes corresponding to the Nb vibrations in the [111] FE *P3m1* structure are respectively, at 155 ($E_u$) and 228 ($A_{2u}$) cm$^{-1}$ in the [111] AFE $P\bar{3}m1$ structure.

All the optical phonon modes are polar in the *Cm* (R) monoclinic structure. While all the A″ modes (Fig. 6) have large IR oscillator strengths and macroscopic electric field along the Cartesian (1, -1, 0) direction, the principal components of the A′ modes involve several directions perpendicular to this direction. The phonon frequencies at all these wavevectors (approaching **q** ~ 0 from these various directions) were computed to obtain the corresponding LO frequencies. Although the absolute LO-TO splitting has decreased in the [111] *Cm* (R) monoclinic structure, there are several polar modes with large oscillator strengths (Fig. 6) and the largest LO-TO splittings in the monoclinic *Cm* (R) structure involve the Nb-O vibrations.

## IV. DISCUSSION

An important aspect of the present study has been to analyze the anisotropy of the Born effective charges to understand the observed enhancement of piezoelectric properties with polarization rotation. The anisotropy of the Born effective charges and their strong orientation dependence contribute to the enhancement of piezoelectric properties with rotation of polarization. As can be seen from Fig. 3, several symmetry lowering directions which maximize the charge contributions from the various atoms can give rise to the observed enhancement. These can explain why the polarization and piezoelectric constants get enhanced in the monoclinic and triclinic structures obtained along these paths. As several possibilities exist, depending on the associated atomic displacements, there can be several monoclinic phases[17-19]. Such phases have been obtained both from experimental[17,18] and theoretical studies[30,59], and it confirms the enhancement of piezoelectric properties with polarization rotation.

The orientation dependence of the piezoelectric and electromechanical properties of PMN-PT (having average tetragonal and local rhombohedral structure) have been measured[60,61], and it is found that the maximum piezoelectric constant and effective electromechanical coupling coefficient are respectively, in directions of 63° and 70.8° from the spontaneous polarization along [111][60]. These orientations are nearly along the pseudocubic crystallographic directions (along the major principal axis of the oxygen atoms) which are consistent with our observations.

First principles calculations[37] of the piezoelectric constants using tetragonal supercells with atomic ordering along [001] reveal significant enhancement of $\varepsilon_{33}$ in going from PZT to PMN-PT due to the large response of the internal coordinates of Pb, Ti, Nb and O atoms to macroscopic strain. These studies[37] also report that Mg atoms contribute little. The present computed Born effective charge tensors for Mg in the lower energy [111] ordered supercell is significantly different from that in the [001] ordered supercell, which can yield additional enhancements to $\varepsilon_{33}$.

Monte Carlo simulations of Pb(Zr$_{1-x}$Ti$_x$)O$_3$ alloys[62] using first principles derived Hamiltonian show that compositional modulation causes the polarization to continuously rotate away from the modulation direction, resulting in monoclinic and triclinic states and huge enhancement of electromechanical response. The orientation dependence of the dipole-dipole interaction[62] in the modulated structure was found to be responsible for these anomalies. The anisotropy of the Born effective charges will also enhance this orientation dependence, in addition to the dipolar energy fluctuations due to compositional modulation[62].

The polar response in PMN is of fundamental interest to understand its spectacular dielectric properties. The phonon instabilities and low frequency polar modes with strong mode effective charges significantly enhance the dielectric properties. Several polar modes are found to couple strongly with the macroscopic field giving rise to giant LO-TO splittings; the splittings have a strong correlation with the phonon instabilities. The largest polar responses are from phonon modes which modulate the covalent bonding of the Nb and O atoms. The anisotropy and significant changes of the Born effective charges with polarization again influence the polar strength leading to distinct vibrational spectra for the [111] and [001] ordered supercells.

## V. CONCLUSIONS

First principles density functional studies have been applied to provide a microscopic understanding of ferroelectric and antiferroelectric states, Born effective charges, electronic dielectric tensor, phonon properties and their variations in 1:2 chemically ordered PMN supercells along the [001] and [111] directions. The comparison between the computed and experimental long wavelength phonon frequencies is very satisfactory, especially as these are idealized supercells, whereas real PMN is disordered. However, there is evidence for small ordered regions[3,63] and our computations would apply directly to understanding the dynamics of such regions. The large polar response in PMN and their variations for ordering along [001] and [111] have an important bearing on their dielectric properties. The multiple phonon instabilities which provide the pathway for polarization rotation shed light on the atomistic mechanisms that govern ferroelectric and piezoelectric enhancements. The principal axis studies of the Born effective charges along with their visualization as charge ellipsoids help understand the important role of the anisotropy in the charge tensors in the observed piezoelectric enhancement with polarization rotation. Detailed analysis of the principal axes of the charge tensors reveal that the maxima of piezoelectric properties are not along the [111] direction of spontaneous polarization, but along the pseudocubic directions which correspond to the major principal axis orientations of the oxygens. A surprising result is that the ground state of a simple well-ordered [111] structure appears to be monoclinic. These studies form an integral part of an ongoing overall program on multiscale modeling of relaxor ferroelectrics, and the present results will help develop models which can study relaxors over various length and time scales.

*Acknowledgements*
This work was supported by the Office of Naval Research under contract number N000149710052. N.C. would like to thank X. Gonze, Bob Downs and A. Asthagiri for helpful discussions. Computations were performed at the Center for Piezoelectrics by Design (CPD), College of William and Mary.

TABLE I: Parameters used in the code OPIUM[45-47] (version beta.3) for generating the norm conserving pseudopotentials.

| Atom | Reference configuration | $r_c$ (bohr) | $q_c(\sqrt{Ryd})$ |
|---|---|---|---|
| Pb | $6s^2, 6p^0, 5d^{10}$ | 1.7, 2.0, 1.75 | 6.05, 5.52, 7.07 |
| Mg | $2s^2, 2p^6, 3d^{0.1}$ | 1.2, 1.5, 1.8 | 9.0, 9.5, 9.0 |
| Nb | $4s^1, 4p^{3.5}, 4d^{3.7}$ | 1.9, 2.0, 2.0 | 6.8, 6.7, 6.8 |
| O | $2s^2, 2p^4$ | 1.16, 1.16 | 9.5, 9.5 |

TABLE II: The space group, relative energies (mRyd/15-atom cell), and lattice vectors for the supercells studied. a'=7.57505 bohr, a''=7.50078 bohr and a'''=7.581 bohr. The *Cm* (E) structure was obtained from the E symmetry unstable phonon mode eigenvector in the FE *P3m1* structure, while the *Cm* (R) structure was obtained via constrained structural relaxation. The ground state has a monoclinic *C2* structure with primitive lattice constants a=10.67688 bohr, b=10.67688 bohr, c=13.08546 bohr, $\alpha$= 90.5518, $\beta$= 89.4482 and $\gamma$=119.0456.

| Ordering direction | Space group | $\Delta E$ LAPW | $\Delta E$ ABINIT | a | b | c |
|---|---|---|---|---|---|---|
| [111] | *P3m1* | 0 | 0 | (-a',a',0) | (a',0,-a') | (-a',-a',-a') |
| [111] | $P\bar{3}m1$ | 12.6 | 11.7 | (-a',a',0) | (a',0,-a') | (-a',-a',-a') |
| [111] | *Cm* (E) |  | -1.11 | (-a',a',0) | (a',0,-a') | (-a',-a',-a') |
| [111] | *Cm* (R) |  | -6.67 | (-a',a',0) | (a',0,-a') | (-a',-a',-a') |
| [111] | *C2* |  | -9.36 |  |  |  |
| [001] | *P4/mmm* | 22.4 | 19.2 | (a'',0,0) | (0,a'',0) | (0,0,3.09a'') |
| [100] | *Pmm2* | 20.3 | 18.4 | (3a''', 0,0) | (0,a''',0) | (0,0,1.01a''') |

TABLE III: Relaxed atomic positions (fractional coordinates) for the AFE $P\bar{3}m1$ structure. The Wyckoff positions *1a*, *2d*, *1b*, *6i* and *3e* have fractional coordinates (0, 0, 0), (1/3, 2/3,z), (0,0,0.5), (x,-x,z) and (0.5,0,0), respectively with ideal positions Pb(*2d*,z)=1/3; Nb(*2d*,z)=5/6; O(*6i*,x)=1/6 and O(*6i*,z)=2/3. The decompositions of the group theoretical irreducible representations (IRD) for the given Wyckoff position at the zone center $\Gamma$ point are also given.

|  | WP | Equivalent positions | Relaxed positions LAPW | Relaxed positions ABINIT | IRD |
|---|---|---|---|---|---|
| Pb | *1a* | (0, 0, 0) |  |  | $A_{2u}+E_u$ |
| Pb | *2d* | ($\frac{1}{3}, \frac{2}{3}$, z), ($\frac{2}{3}, \frac{1}{3}$,-z) | .3370 (z) | .3355 (z) | $A_{1g}+A_{2u}+E_u+E_g$ |
| Mg | *1b* | (0,0,0.5) |  |  | $A_{2u}+E_u$ |
| Nb | *2d* | ($\frac{1}{3}, \frac{2}{3}$, z), ($\frac{2}{3}, \frac{1}{3}$, -z) | .8215 (z) | .8206 (z) | $A_{1g}+A_{2u}+E_u+E_g$ |
| O | *6i* | (x,-x,z),(x,2x,z),(-2x,-x,z), (-x,x,-z),(2x,x,-z),(-x,-2x,-z) | .1694 (x) .6725 (z) | .1703 (x) .6730 (z) | $2A_{1g}+A_{1u}+A_{2g}+2A_{2u}+3E_u+3E_g$ |
| O | *3e* | (0.5, 0, 0) |  |  | $A_{1u}+2A_{2u}+3E_u$ |

TABLE IV: Relaxed FE *P3m1* structure. The Wyckoff positions *1a*, *1b*, *1c*, and *3d* have fractional coordinates (0,0,z), (1/3,2/3,z), (2/3,1/3,z) and (x,-x,z), respectively.

|    | WP | Equivalent positions | Fractional coordinates LAPW | Fractional coordinates ABINIT | IRD[†] |
|----|----|---------------------|----------------------------|------------------------------|--------|
| Pb | 1a | (0,0,z) | .0306 (z) | .0272(z) | $A_1+E$ |
| Pb | 1b | $(\frac{1}{3},\frac{2}{3},z)$ | .3772 (z) | .3752(z) | $A_1+E$ |
| Pb | 1c | $(\frac{2}{3},\frac{1}{3},z)$ | .6922 (z) | .6940(z) | $A_1+E$ |
| Mg | 1a | (0,0,z) | .5   (z) | .5   (z) | $A_1+E$ |
| Nb | 1b | $(\frac{1}{3},\frac{2}{3},z)$ | .8238 (z) | .8211(z) | $A_1+E$ |
| Nb | 1c | $(\frac{2}{3},\frac{1}{3},z)$ | .1765 (z) | .1766(z) | $A_1+E$ |
| O  | 3d | (x,-x,z), (x,2x,z), (-2x,-x,z) | .1751 (x) .6527 (z) | .1735(x) .6514(z) | $2A_1+A_2+3E$ |
| O  | 3d | (x,-x,z), (x,2x,z), (-2x,-x,z) | .3299 (x) .3076 (z) | .3309(x) .3047(z) | $2A_1+A_2+3E$ |
| O  | 3d | (x,-x,z), (x,2x,z), (-2x,-x,z) | .5050 (x) -.0203 (z) | .5043(x) -.0229(z) | $2A_1+A_2+3E$ |

[†] Decomposition of the group theoretical irreducible representation at the $\Gamma$ point

TABLE V: Relaxed AFE *P4/mmm* structure. The Wyckoff positions *1a*, *2g*, *1d*, *2h*, *1c*, *2e* and *4i* have fractional coordinates (0,0,0), (0,0,z), (0.5,0.5,0.5), (0.5,0.5,z), (0.5,0.5,0), (0,0.5,0.5), (0.5,0.5,z) and (0,0.5,z) respectively, and the ideal positions have z values of 1/3, 1/6, 1/3 and 1/6 for the Pb(*2g*), Nb(*2h*), O(*2h*) and O(*4i*).

|    | WP | Related positions | Relaxed positions LAPW | Relaxed positions ABINIT | IRD[†] |
|----|----|------------------|------------------------|--------------------------|--------|
| Pb | 1a | (0,0,0) | | | $A_{2u}+E_u$ |
| Pb | 2g | (0,0,z), (0,0,-z) | .3699 (z) | .3694 (z) | $A_{1g}+A_{2u}+E_u+E_g$ |
| Mg | 1d | (0.5,0.5,0.5) | | | $A_{2u}+E_u$ |
| Nb | 2h | (0.5,0.5,z), (0.5,0.5,-z) | .1698 (z) | .1697 (z) | $A_{1g}+A_{2u}+E_u+E_g$ |
| O  | 1c | (0.5,0.5,0) | | | $A_{2u}+E_u$ |
| O  | 2e | (0,0.5,.5),(0.5,0,0.5) | | | $A_{2u}+B_{2u}+2E_u$ |
| O  | 2h | (0.5,0.5,z),(0.5,0.5,-z) | .3217 (z) | .3199 (z) | $A_{1g}+A_{2u}+E_g+E_u$ |
| O  | 4i | (0,0.5,z),(0.5,0,z), (0,0.5,-z),(0.5,0,-z) | .1561 (z) | .1569 (z) | $A_{1g}+A_{2u}+B_{1g}+B_{2u}+2E_u+2E_g$ |

[†]Decomposition of the group theoretical irreducible representation at the $\Gamma$ point

TABLE VI: The computed monoclinic $Cm$ (E) and $Cm$ (R) structures involving polarization rotation in PMN. The $Cm$ (E) structure was obtained from the 45i cm$^{-1}$ E unstable TO phonon mode eigenvector in the FE $P3m1$ structure, while the $Cm$ (R) structure was obtained via constrained structural relaxation. The cell constants are identical to the FE $P3m$1 structure and are given in Table II.

|  | x | y | z |
|---|---|---|---|
| $Cm$ (E) | | | |
| Pb | -.0083 | -.0166 | .0288 |
| Pb | .3389 | .6778 | .3768 |
| Pb | .6578 | .3156 | .6956 |
| Mg | .0041 | .0082 | .5016 |
| Nb | .3355 | .6710 | .8227 |
| Nb | .6700 | .3400 | .1782 |
| O1 | .1955 | .8528 | .6520 |
| O2 | .1827 | .3653 | .6551 |
| O3 | .6573 | .8528 | .6520 |
| O4 | .3403 | .1828 | .3079 |
| O5 | .8425 | .1828 | .3079 |
| O6 | .8416 | .6832 | .3032 |
| O7 | .5203 | .0407 | -.0247 |
| O8 | .0075 | .5308 | -.0196 |
| O9 | .5233 | .5308 | -.0196 |
| $Cm$ (R) | | | |
| Pb | -.0214 | -.0428 | -.0008 |
| Pb | .3145 | .6290 | .3374 |
| Pb | .6411 | .2822 | .6747 |
| Mg | .0004 | .0008 | .4979 |
| Nb | .3330 | .6660 | .8179 |
| Nb | .6696 | .3392 | .1810 |
| O1 | .1988 | .8709 | .6736 |
| O2 | .1845 | .3690 | .6562 |
| O3 | .6721 | .8709 | .6736 |
| O4 | .3533 | .2084 | .3274 |
| O5 | .8551 | .2084 | .3274 |
| O6 | .8467 | .6934 | .3054 |
| O7 | .5236 | .0472 | -.0079 |
| O8 | .0210 | .5415 | -.0054 |
| O9 | .5206 | .5415 | -.0054 |

TABLE VII: The computed charge tensors for the [111] FE $P3m1$ structure. The principal axes are along the Cartesian [1$\bar{1}$0], [11$\bar{2}$] and [111] directions for all the Pb, Mg and Nb atoms respectively. The oxygen atoms have their major principal axes (nearly) along the pseudocubic directions.

|  | WP | Principal value 1 | Principal value 2 | Principal value 3 |
|---|---|---|---|---|
| Pb | 1a | 4.16 | 4.16 | 3.06 |
| Pb | 1b | 3.09 | 3.09 | 4.54 |
| Pb | 1c | 3.80 | 3.80 | 3.25 |
| Mg | 1a | 2.86 | 2.86 | 2.83 |
| Nb | 1b | 6.87 | 6.87 | 6.41 |
| Nb | 1c | 6.83 | 6.83 | 5.91 |
| O | 3d | -2.56 | -2.38 | -3.59 |
| O | 3d | -2.06 | -2.41 | -4.14 |
| O | 3d | -2.18 | -2.22 | -5.52 |

TABLE VIII: The computed dynamic charge tensors for the [111] AFE $P\bar{3}m1$ structure. The principal axes directions are as given for the FE $P3m1$ structure in Table VII.

|    | WP | Principal value 1 | Principal value 2 | Principal value 3 |
|----|----|-------------------|-------------------|-------------------|
| Pb | 1a | 4.34 | 4.34 | 3.04 |
| Pb | 2d | 3.76 | 3.76 | 4.39 |
| Mg | 1b | 2.82 | 2.82 | 2.92 |
| Nb | 2d | 6.95 | 6.95 | 6.32 |
| O  | 6i | -2.49 | -2.52 | -3.92 |
| O  | 3e | -2.41 | -2.32 | -5.59 |

TABLE IX: The computed dynamic charge tensors for the [001] $P4/mmm$ structure obtained in the present study and its comparisons with the reported charge tensors for the [001]$_{NNM}$ supercell of PMN by Prosandeev et al.[30], and average charge tensor values reported[37] for PMN-PT using tetragonal [001] ordered supercells.

|    | WP | $Z^*_{xx}$ | $Z^*_{yy}$ | $Z^*_{zz}$ | $Z^{*\ 30}_{xx}$ | $Z^{*\ 30}_{yy}$ | $Z^{*\ 30}_{zz}$ | $Z^{*37}$ |
|----|----|------|------|------|------|------|------|------|
| Pb | 1a | 3.92 | 3.92 | 3.41 | 3.90 | 3.90 | 3.33 | 3.6 |
| Pb | 2g | 3.81 | 3.81 | 3.53 | 3.8  | 3.8  | 3.42 |     |
| Mg | 1d | 2.01 | 2.01 | 2.42 | 1.9  | 1.9  | 2.56 | 1.9 |
| Nb | 2h | 8.38 | 8.38 | 7.74 | 8.76 | 8.76 | 7.64 | 7.4 |
| O  | 1c | -2.76 | -2.76 | -5.26 | -2.75 | -2.75 | -5.18 | -2.1, -5.1 |
| O  | 2e | -3.99 | -1.87 | -2.85 | -3.99 | -1.76 | -2.91 |     |
| O  | 2h | -2.18 | -2.18 | -4.78 | -2.15 | -2.15 | -4.62 |     |
| O  | 4i | -2.15 | -6.51 | -1.96 | -2.18 | -6.84 | -1.94 |     |

TABLE X: The computed dynamic charge tensors for the [111] FE monoclinic $Cm$ (R) structure. O1 and O3, O5, and O6, O8 and O9 are crystallographically equivalent.

|    | Principal value 1 | Principal value 2 | Principal value 3 |
|----|-------------------|-------------------|-------------------|
| Pb | 2.85 | 3.71 | 4.11 |
| Pb | 3.07 | 3.89 | 4.36 |
| Pb | 3.05 | 3.60 | 3.96 |
| Mg | 2.78 | 2.86 | 3.00 |
| Nb | 5.82 | 6.48 | 6.62 |
| Nb | 5.73 | 6.49 | 6.52 |
| O1 | -3.59 | -2.60 | -2.29 |
| O2 | -4.12 | -2.25 | -1.96 |
| O4 | -4.11 | -2.27 | -2.08 |
| O5 | -3.59 | -2.49 | -2.36 |
| O7 | -5.01 | -2.25 | -2.01 |
| O8 | -5.22 | -2.33 | -1.96 |

TABLE XI: The computed polar longitudinal optic (LO) and transverse optic (TO) phonon frequencies and mode effective charges of [111] FE *P3m1* structure. The optically inactive and non-polar $A_2$ phonon modes have frequencies of 27, 98 and 200 cm$^{-1}$ respectively. LO1 and LO2 are the LO frequencies obtained from linear response studies using electric field perturbation and the infrared oscillator strengths respectively. The correspondence between the LO1 and TO modes is determined from the correlation matrix and the LO1 frequencies listed correspond to the phonon modes having the largest overlap.

|       | TO cm$^{-1}$ | Mode effective charge | LO1 cm$^{-1}$ | LO2 cm$^{-1}$ |
|-------|------|-------|----------|------|
| E     |      |       |          |      |
|       | 45i  | 7.89  | 107, 404 | 261  |
|       | 27   | 2.49  | 35       | 52   |
|       | 49   | 5.58  | 107, 404 | 154  |
|       | 124  | 1.42  | 129      | 140  |
|       | 143  | 1.11  | 145      | 151  |
|       | 180  | 8.71  | 403      | 353  |
|       | 250  | 1.01  | 261      | 254  |
|       | 272  | 3.43  | 317      | 301  |
|       | 324  | 2.14  | 404      | 334  |
|       | 327  | 0.01  | 327      | 327  |
|       | 361  | 0.27  | 361      | 361  |
|       | 502  | 4.06  | 563      | 540  |
|       | 544  | 1.08  | 563      | 546  |
|       | 629  | 4.07  | 701      | 657  |
| $A_1$ |      |       |          |      |
|       | 66   | 1.01  | 67       | 69   |
|       | 69   | 3.36  | 71       | 85   |
|       | 121  | 5.98  | 153, 442 | 232  |
|       | 254  | 6.22  | 442      | 321  |
|       | 259  | 8.89  | 442, 667 | 421  |
|       | 320  | 0.49  | 320      | 320  |
|       | 405  | 0.45  | 406      | 405  |
|       | 411  | 1.19  | 442      | 415  |
|       | 564  | 5.19  | 667      | 618  |
|       | 776  | 0.16  | 776      | 776  |
|       | 811  | 0.46  | 811      | 811  |

TABLE XII: Computed long wavelength polar phonon frequencies and mode assignments in the AFE $P\bar{3}m1$ structure (LO notations as in table XI). The non-polar modes are $A_{1g}$ (19, 250, 381 and 754 cm$^{-1}$), $A_{1u}$ (51i cm$^{-1}$), $A_{2g}$ (65i, 181 cm$^{-1}$) and the doubly degenerate $E_g$ modes (41i, 59, 257, 320 and 564 cm$^{-1}$).

|  | TO cm$^{-1}$ | Mode effective charge | LO1 cm$^{-1}$ | LO2 cm$^{-1}$ |
|---|---|---|---|---|
| $E_u$ | | | | |
|  | 110i | 8.78 | 94, 397 | 369 |
|  | 68i | 0.12 | 68i | 68i |
|  | 38 | 0.32 | 38 | 38 |
|  | 155 | 7.15 | 183, 397 | 285 |
|  | 192 | 3.73 | 326 | 250 |
|  | 256 | 1.95 | 315 | 257 |
|  | 350 | 0.78 | 397 | 352 |
|  | 494 | 3.71 | 539 | 523 |
|  | 616 | 4.14 | 685 | 645 |
| $A_{2u}$ | | | | |
|  | 90i | 7.67 | 113, 409 | 292 |
|  | 17 | 1.28 | 17 | 25 |
|  | 228 | 9.50 | 409 | 385 |
|  | 322 | 1.63 | 377 | 328 |
|  | 387 | 1.07 | 409 | 390 |
|  | 552 | 5.82 | 665 | 613 |
|  | 806 | 0.56 | 808 | 807 |

TABLE XIII: Computed long wavelength polar phonon frequencies and mode assignments in the *P4/mmm* structure (LO notations as in table XI). The non polar modes are $A_{1g}$ (135, 172, 420 and 784 cm$^{-1}$), $B_{1g}$ (276 cm$^{-1}$), $B_{2u}$ (239, 260 cm$^{-1}$) and $E_g$ (63i, 43, 127, 226, 471 cm$^{-1}$).

|       | TO cm$^{-1}$ | Mode effective charge | LO1 cm$^{-1}$ | LO2 cm$^{-1}$ |
|-------|------|-----------|------|------|
| $E_u$ |      |           |      |      |
|       | 156i | 12.02     | 665  | 531  |
|       | 7.3i | 3.93      | 58   | 67   |
|       | 61   | 2.44      | 128  | 77   |
|       | 130  | .521      | 130  | 132  |
|       | 216  | 1.01      | 227  | 220  |
|       | 273  | .358      | 272  | 273  |
|       | 331  | 3.41      | 403  | 361  |
|       | 487  | 5.23      | 580  | 545  |
|       | 593  | 1.82      | 666  | 599  |
| $A_{2u}$ |   |           |      |      |
|       | 54   | 6.43      | 195  | 184  |
|       | 103  | 5.14      | 124  | 162  |
|       | 188  | 3.56      | 260  | 232  |
|       | 238  | 9.86      | 447  | 454  |
|       | 411  | 1.48      | 419  | 416  |
|       | 566  | 5.23      | 646  | 614  |
|       | 742  | 3.23      | 778  | 757  |

TABLE XIV: Computed long wavelength phonon frequencies and mode assignments in the [111] FE *Cm* (R) monoclinic structure (LO notations as in table XI). The monoclinic unique axis is along the Cartesian (1, -1, 0) direction (hexagonal **a**-axis), and the A″ (A′) phonon modes have their macroscopic field parallel (perpendicular) to this direction.

|     | TO cm$^{-1}$ | Mode effective charge | LO1 cm$^{-1}$ | LO2 cm$^{-1}$ |
|---|---|---|---|---|
| A″ |  |  |  |  |
|  | 42i | 5.8 | 121, 405 | 174 |
|  | 62 | 5.5 | 121, 405 | 170 |
|  | 77 | 0.7 | 81 | 83 |
|  | 98 | 0.8 | 99 | 106 |
|  | 146 | 0.2 | 146 | 146 |
|  | 169 | 3.2 | 204 | 222 |
|  | 204 | 3.4 | 245 | 256 |
|  | 217 | 5.0 | 271 | 310 |
|  | 255 | 3.5 | 270 | 288 |
|  | 275 | 2.2 | 293 | 294 |
|  | 322 | 2.1 | 332 | 333 |
|  | 330 | 0.6 | 330 | 331 |
|  | 354 | 0.7 | 353 | 355 |
|  | 515 | 4.6 | 576 | 561 |
|  | 531 | 0.5 | 531 | 532 |
|  | 642 | 3.9 | 708 | 667 |
| A′ |  |  |  |  |
|  | 30 | 5.8 | 119, 161 | 124 |
|  | 47 | 0.8 | 47 | 49 |
|  | 67 | 5.6 | 160 | 178 |
|  | 69 | 1.5 | 69 | 74 |
|  | 81 | 3.9 | 118 | 108 |
|  | 126 | 4.1 | 149 | 192 |
|  | 139 | 1.4 | 149 | 154 |
|  | 147 | 2.3 | 160 | 174 |
|  | 227 | 6.7 | 425 | 339 |
|  | 251 | 4.6 | 379 | 315 |
|  | 257 | 7.3 | 425 | 371 |
|  | 267 | 2.6 | 282 | 289 |
|  | 276 | 6.9 | 416 | 390 |
|  | 317 | 1.2 | 317 | 320 |
|  | 325 | 1.2 | 330 | 329 |
|  | 350 | 0.1 | 350 | 350 |
|  | 368 | 0.3 | 368 | 368 |
|  | 387 | 1.0 | 387 | 390 |
|  | 402 | 0.7 | 402 | 403 |
|  | 536 | 4.1 | 560 | 571 |
|  | 552 | 4.1 | 584 | 587 |
|  | 597 | 4.3 | 604 | 631 |
|  | 632 | 4.0 | 697 | 660 |
|  | 797 | 0.9 | 797 | 798 |
|  | 810 | 0.6 | 811 | 811 |

FIG. 1: The polyhedral representation of the supercells for [111] FE and [001] AFE structures. The coordination polyhedra around the Mg (light) and Nb (dark) are shown. The Pb atoms are shown as spheres (distinct shade and sizes represent different Wyckoff positions).

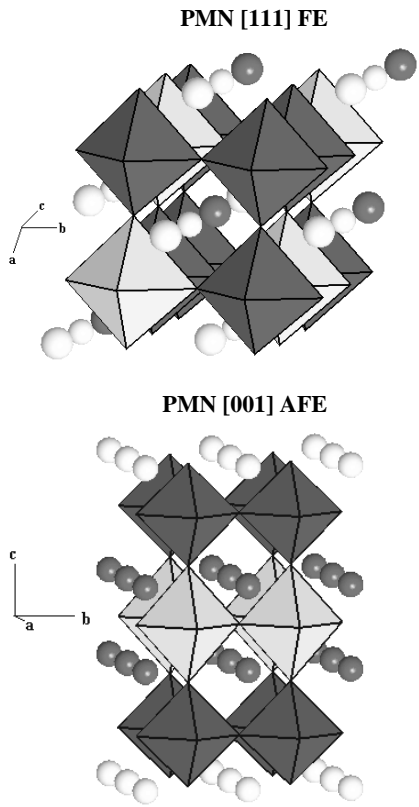

FIG. 2: Frozen phonon energy surfaces for various ordered supercells. (a) The changes in total energy involving structural changes from $P\bar{3}m1$ AFE to $P3m1$ FE obtained using all electron LAPW and pseudopotential ABINIT. In (b) and (c) the double wells corresponding to the lattice instabilities from $P3m1$ to $Cm$ (E) and from $Cm$ (E) to $Cm$ (R) are shown.

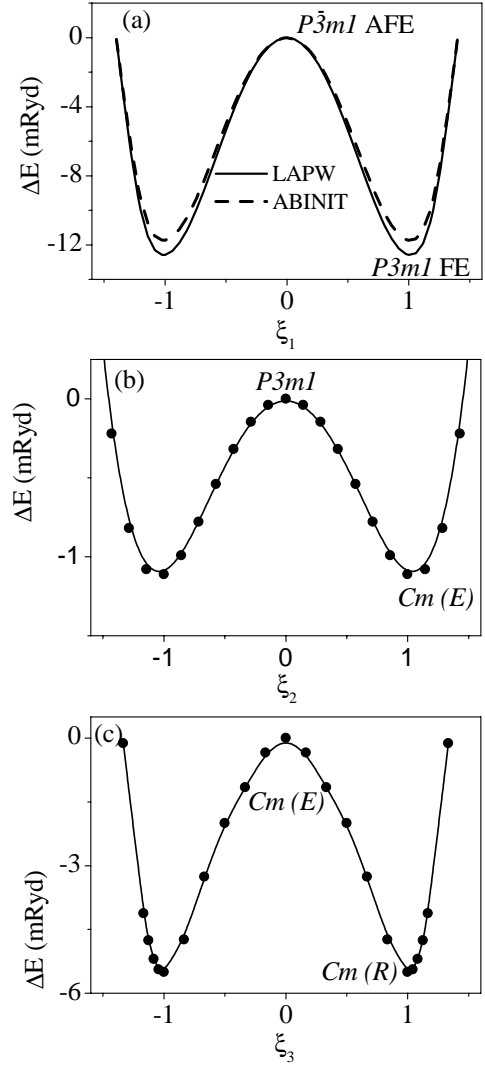

Fig. 3: Representation quadric of the Born effective charge tensor as a "charge ellipsoid" generated from the software xtaldraw[50]. (a) and (c) denote the [111] FE structures, and (b) and (d) are [111] AFE . (a-b) give the view with the hexagonal *c*-axis as vertical, and viewed down the *a*–axis, while (c-d) show the view in the horizontal *ab*-plane. The lightest shade corresponds to Pb atoms, and the Mg atoms are located midway between the Pb atoms.

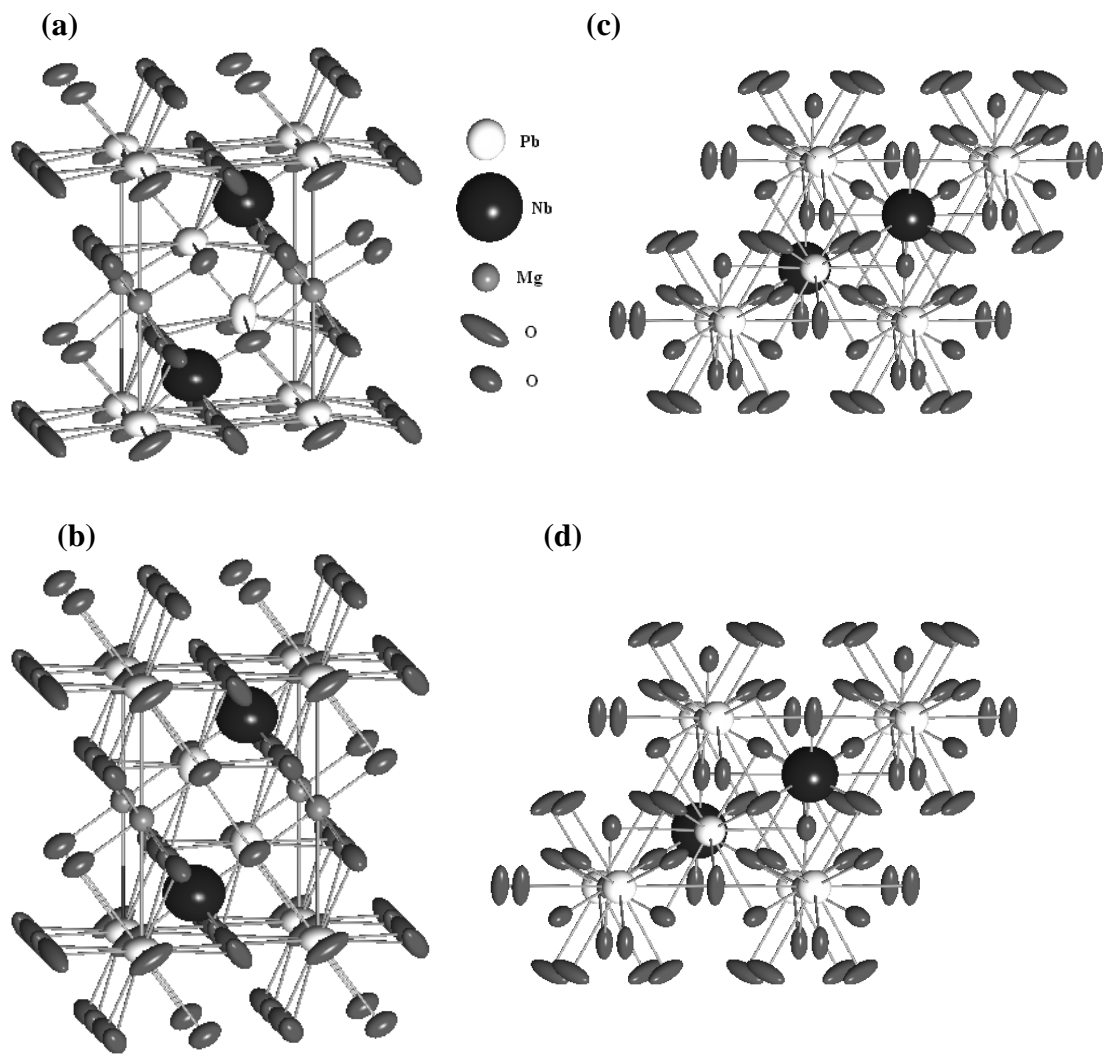

Fig. 4: (a) The computed zone center phonon frequencies of various ordered supercells and their comparisons with available Raman (star[30], triangle[8]) and infrared[30] (hexagonal TO, triangle LO) data. (b) The polar phonon frequencies and mode assignments in [111] FE structure. The $A_1$ and E modes are both Raman and infrared active. LO1 gives the LO frequencies using the IR oscillator strength[43,44] approach, and LO2 and LO3 give the LO frequencies obtained from linear response of electric field perturbations. LO3 involves symmetrized charge tensors for the oxygens.

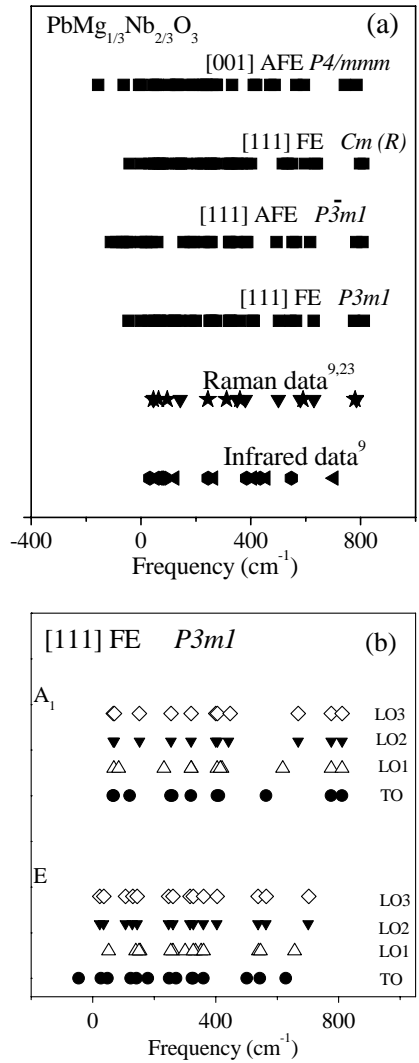

Fig. 5: The long wavelength frequency distribution (Total) and the displacement weighted spectra giving the dynamical contributions from various atoms in the *Cm* (R) (a-e), *P3m1* (f-j), *P$\bar{3}$m1* (k-o) and *P4/mmm* (p-t) structures. These provide a graphical representation of the long wavelength phonon eigenvectors and show that despite the complex structure of PMN, the vibrational contributions from various cations are quite well separated.

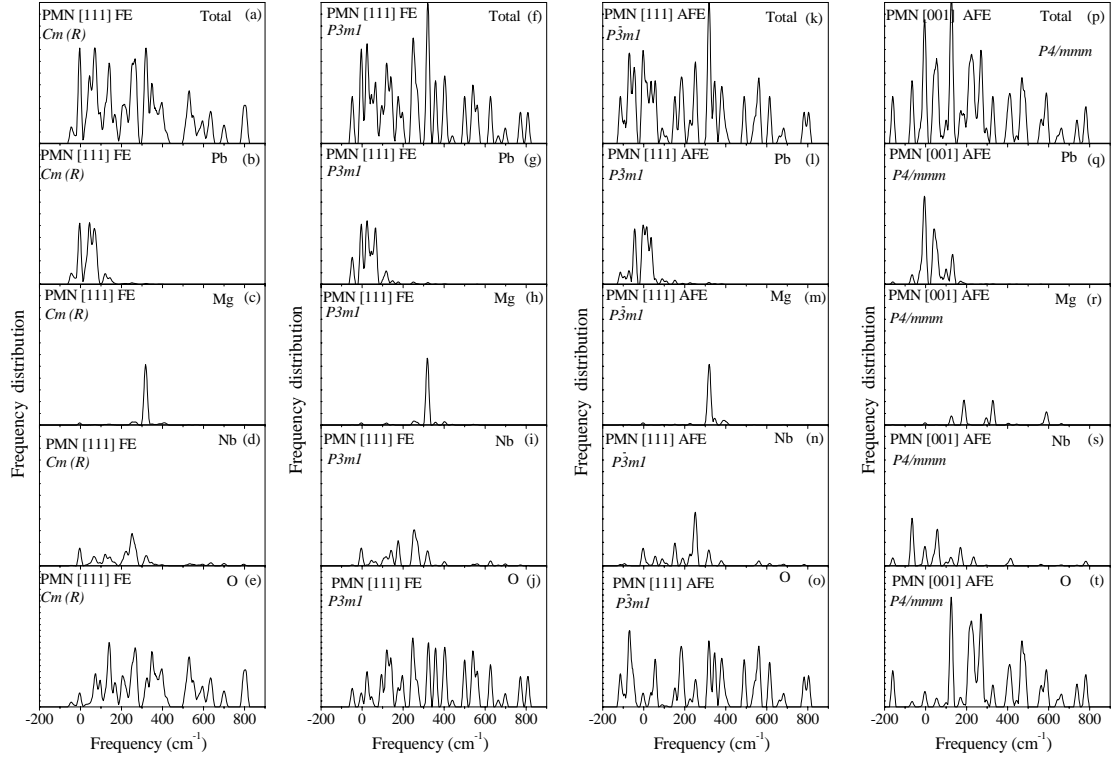

Fig. 6: (a) The computed infrared oscillator strengths (atomic units: 1 a.u.=253.2638413 m$^3$/s$^2$) resolved along (perpendicular to) the tetragonal axes corresponding to phonon modes with A$_{2u}$ (E$_u$) symmetry in the *P4/mmm* structure. (b) and (c) The IR oscillator strengths along (perpendicular to) hexagonal axes as dashed (full) lines, and they represent the A$_1$ (E) and A$_{2u}$ (E$_u$) types polar phonon modes in the *P3m1* and *P$\bar{3}$m1* structures respectively. (d) The oscillator strengths along (perpendicular to) the monoclinic axis in the *Cm* (R) structure corresponding to the A$''$ (A$'$) modes. The associated TO phonon frequencies (cm$^{-1}$) are indicated on top.

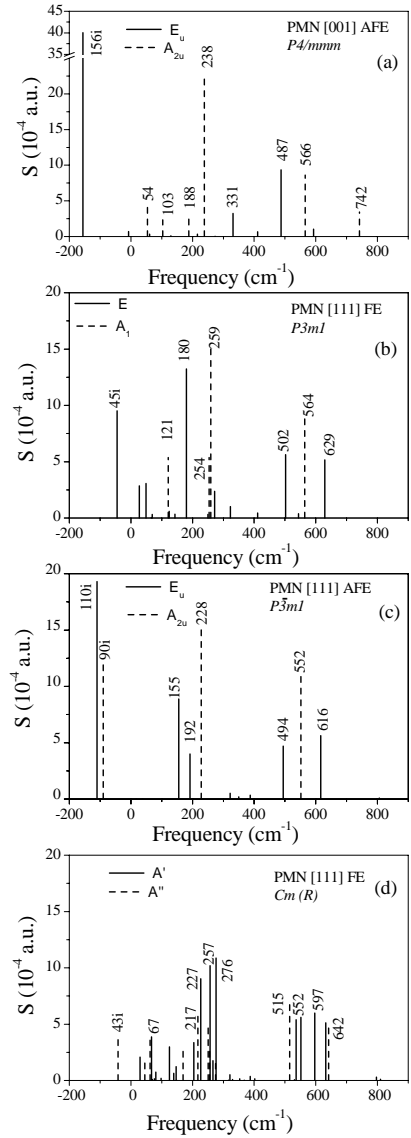